\documentclass[a4paper,fleqn]{cas-dc}
\usepackage{subfig}
\usepackage[authoryear,longnamesfirst]{natbib}
\usepackage{amsmath}
\usepackage{tikz}
\usepackage{xcolor}
\usetikzlibrary{arrows.meta,positioning,fit}
\providecommand{\arcsec}{\ensuremath{^{\prime\prime}}}
\providecommand{\degr}{\ensuremath{^{\circ}}}

\begin{document}
\let\WriteBookmarks\relax
\def\floatpagepagefraction{1}
\def\textpagefraction{.001}

\shorttitle{CosmoSim on real gravitational lenses}
\shortauthors{Normann et~al.}

\title[mode=title]{CosmoSim: Lens-parameter estimation of the Einstein Cross NGC~6505 using machine learning}

\author[1]{Ben David Normann}
\cormark[1]
\ead{ben.d.normann@ntnu.no}
\credit{Conceptualization, Supervision, Writing -- review and editing}
\affiliation[1]{organization={Department of ICT and Natural Sciences, Norwegian University of Science and Technology (NTNU)},
            city={{\AA}lesund},
            country={Norway}}

\author[2]{Oda Hjemli}
\ead{oda.hjemli@ntnu.no}
\credit{Software, Investigation, Data curation, Writing -- original draft}
\affiliation[2]{organization={Department of Physics, Norwegian University of Science and Technology (NTNU)},
            city={Trondheim},
            country={Norway}}

\author[1]{Hans Georg Schaathun}
\ead{hasc@ntnu.no}
\credit{Software, Supervision, Writing -- review and editing}

\cortext[1]{Corresponding author}

\begin{abstract}
   CosmoSim is an open-source tool for simulating gravitational lenses and
   reconstructing lens mass distributions, either by ray tracing a parametric
   lens model or through a Roulette expansion of the lens potential. It is
   also used to generate the large, labelled synthetic image sets needed to
   train machine-learning (ML) parameter estimators. We present the first
   application of CosmoSim to real data. A convolutional neural network
   trained only on CosmoSim images estimates Singular Isothermal Ellipsoid
   lens and source parameters directly from observed images. We test the
   pipeline on 21 images of 19 strong lenses from Euclid and the AGEL survey,
   after removing the foreground lens light with a custom decontamination
   tool built on AstroPy. As a case study we reconstruct the Einstein Cross
   NGC~6505 both from the lens model published by the discovery team and
   from parameters predicted by our ML model. Both reconstructions recover
   the observed four-image structure, and the ML-based reconstruction
   performs at least as well as the literature-based one: it scores better
   on masked structural similarity, intersection-over-union and Chamfer
   distance, with an essentially equal Hausdorff distance. Across the
   ensemble, the predicted Einstein radii are consistent with catalogued
   values within the quoted uncertainties (mean relative difference 0.08 for
   Euclid and 0.19 for AGEL lenses), and the enclosed masses agree to within
   a median of 0.12~dex. The predicted ellipticities differ systematically
   from the catalogued values, while the comparison of orientations is
   inconclusive. These results establish CosmoSim as a tool that can be
   applied directly to observational data.
\end{abstract}

\begin{highlights}
\item First application of CosmoSim to observed strong gravitational lenses
\item CNN trained only on CosmoSim simulations estimates SIE lens parameters
\item ML reconstruction of NGC 6505 matches one built from the published model
\item Predicted Einstein radii agree with Euclid and AGEL catalogue values
\item Predicted ellipticities differ systematically from catalogue values
\end{highlights}


\maketitle

\section{Introduction}
\label{sec:intro}

Gravitational lensing is one of the few direct probes of the total
mass distribution of galaxies, including their dark matter content, and allows for a variety of applications, such as the estimation of cosmological probes like the Hubble constant or the subhalo mass
function. The next generation of wide-field space surveys, in particular
Euclid and the forthcoming Chinese Survey Space Telescope (CSST), are
expected to increase the number of known strong lens systems by two to
three orders of magnitude \citep{EuclidSLDE2025}. Exploiting this sample
requires lens-modelling software that is both fast enough to run at scale
and flexible enough to be trained, or retrained, on large synthetic
datasets.

CosmoSim is a tool developed for exactly this purpose: visualising
gravitational lenses and reconstructing lens mass distributions, either by
direct ray tracing of a parametric lens model or through the parameter-free
Roulette expansion of the lens potential
\citep{Clarkson2016a,Clarkson2016b}, as implemented in CosmoSim
\citep{Schaathun2023}. Because it is fast and fully
labelled, CosmoSim is also used to generate the large synthetic image sets
needed to train machine-learning (ML) parameter estimators. CosmoSim has been applied successfully to lens-mass reconstruction
\citep{hgs2023nik,Schaathun2023}, and a forthcoming software-release paper
will describe the tool in detail. These previous studies have, however, been
limited to synthetic data; in the present paper we apply CosmoSim to
lens-mass reconstruction with real data for the first time. This is done with an ML model
trained on CosmoSim's synthetic output, and serves as a demonstration that the pipeline works as intended.

As a headline result we reconstruct the Einstein Cross NGC~6505, discovered in
Euclid imaging \citep{ORiordan2025}, directly from an observational image,
and we quantify how this reconstruction compares with the lens model
published by the discovery team. We also test the ML model on strong lenses from the AGEL survey and the
Euclid Strong Lensing Discovery Engine, and quantify how well it performs on
these systems. 

The remainder of this paper is structured as follows: Section \ref{sec:theory} describes the relevant theoretical background for gravitational lens mass reconstruction in general. Section
\ref{sec:method} describes the main tool and how it is used to train the ML parameter
estimator on synthetic data. The full processing pipeline is also described. Section \ref{sec:results} reports the results: how the resulting reconstruction of NGC~6505 performs, in
particular relative to the discovery team's own lens model, together with
the validation results over the AGEL and Euclid ensemble. Finally, in Section
\ref{sec:conclusion} we conclude. The underlying theoretical background and
the full analysis are given in the associated thesis
\citep{Hjemli2026}, on which this paper is based. Appendix~\ref{app:ensemble}
gives the full set of decontaminated/reconstructed image pairs for the
observational ensemble.

\section{Lens-parameter reconstruction}
\label{sec:theory}
The theory behind gravitational lensing is well established, and we refer the reader to \citet{Schneider1992}, or newer references, such as \citet{bok:principlesOfGravLens18} for a comprehensive review. In this section we summarise the key concepts and equations relevant to our analysis.

\subsection{Lens model: the singular isothermal ellipsoid}
\label{sec:sie}
A source at angular position $\vec\beta$ is mapped to image positions
$\vec\theta$ by the lens equation
\begin{equation}
\vec\beta = \vec\theta - \nabla\psi(\vec\theta), \qquad \nabla^2\psi = 2\kappa,
\label{eq:lenseq}
\end{equation}
where $\psi$ is the scaled projected lens potential and
$\kappa=\Sigma/\Sigma_\text{cr}$ the convergence. For an isothermal sphere
(SIS) with velocity dispersion $\sigma_\text{v}$, the surface density falls off as
$\Sigma\propto 1/|\vec\theta|$, giving $\psi = \theta_\text{E}|\vec\theta|$ with
Einstein radius
\begin{equation}
\theta_\text{E} = 4\pi\left(\frac{\sigma_\text{v}}{c}\right)^2\frac{D_\text{LS}}{D_\text{S}}.
\label{eq:thetaE}
\end{equation}
Real lens galaxies and their dark-matter haloes are, however, elliptical,
and the isothermal profile of the combined mass is well described by the
singular isothermal ellipsoid (SIE) \citep{Kormann1994}. In lens-centred
coordinates $(\theta_1,\theta_2)$, rotated counter-clockwise by the
orientation angle $\phi_\text{L}$ from the image $x$-axis, the circular
radius is replaced by the elliptical radius
$\zeta=\sqrt{\theta_1^2+\epsilon^2\theta_2^2}$, where the ellipticity
$0<\epsilon\le1$ is the minor-to-major axis ratio. Since $\epsilon\le1$, the
major axis of the mass distribution lies along $\theta_2$, i.e.\ at
$\phi_\text{L}+90^\circ$. The convergence and potential become
\begin{align}
\kappa(\vec\theta) &= \frac{\sqrt{\epsilon}}{2}\,\frac{\theta_\text{E}}{\zeta}, \label{eq:sie_kappa}\\
\psi(\vec\theta) &= \theta_\text{E}\frac{\sqrt{\epsilon}}{\epsilon'}\,|\vec\theta|\,
\Big[\sin\varphi\,\arcsin\!\left(\epsilon'\sin\varphi\right) \nonumber\\
&\qquad +\cos\varphi\,\operatorname{arsinh}\!\big(\tfrac{\epsilon'}{\epsilon}\cos\varphi\big)\Big],
\label{eq:sie_psi}
\end{align}
with $\epsilon'=\sqrt{1-\epsilon^2}$ and $\varphi$ the polar angle. The SIE reduces to
the SIS for $\epsilon=1$. Unlike the SIS, it can produce four images of a single
source, as observed in Einstein crosses such as NGC~6505. The lens is thus
described by three parameters: $\theta_\text{E}$, the ellipticity
$\epsilon$ and the orientation $\phi_\text{L}$ (Table~\ref{tbl:trainrange}).

\subsection{Source model: the S\'ersic light profile}
\label{sec:sersic}
Lensing conserves surface brightness, so the lensed image is
$I(\vec\theta)=I_\text{S}(\vec\beta(\vec\theta))$, where $I_\text{S}$ is the source light
profile. For the source we use the S\'ersic profile \citep{Sersic1963},
which describes the light of a range of different galaxies:
\begin{equation}
I_\text{S}(R) = I_\text{e} \exp\!\left\{-b_n\left[\left(\frac{R}{R_\text{Eff}}\right)^{1/n}-1\right]\right\},
\label{eq:sersic}
\end{equation}
where $R$ is the distance from the source centre, $R_\text{Eff}$ the
effective (half-light) radius, $I_\text{e}$ the intensity at $R_\text{Eff}$ and
$n$ the S\'ersic index, which sets how steeply the light falls off ($n=1$
gives an exponential disc, $n=4$ a de~Vaucouleurs profile). The constant
$b_n$ is fixed by requiring that $R_\text{Eff}$ enclose half the total
light, $\gamma(2n,b_n)=\Gamma(2n)/2$, and is well approximated by
$b_n\approx1.9992n-0.3271$ \citep{GrahamDriver2005}. Integrating
Eq.~(\ref{eq:sersic}) gives the total flux
\begin{equation}
\begin{gathered}
F = 2\pi n\, I_\text{e} R_\text{Eff}^2\,\frac{e^{b_n}}{b_n^{2n}}\,\Gamma(2n),\\
m_\text{app} = -2.5\log_{10}F + \text{ZP},
\end{gathered}
\label{eq:sersicflux}
\end{equation}
where ZP is the zero-point magnitude of the telescope and filter. In
CosmoSim the source can therefore be specified by observable quantities --
the apparent magnitude $m_\text{app}$, $R_\text{Eff}$ and $n$ -- with
$I_\text{e}$ following from Eq.~(\ref{eq:sersicflux}). Alternatively, the
brightness can be set by a dimensionless total luminosity
$L_\text{S\'ersic}$, which is what we use for the training data in this
paper. Together with the source position $(R_\text{S},\phi_\text{S})$ this
completes the parameter set in Table~\ref{tbl:trainrange}.
\section{Tools and method}
\label{sec:method}
To actually train an ML model for lens parameter reconstruction, we use the recently developed CosmoSim toolbox. All simulations in this
paper were made with the development branch \texttt{physical implementation}
of CosmoSim in April and May 2026; the nearest archived release is v2.8.0
of 12~April 2026~\citep{cosmosim2.7}. The training data (6~April) and the
reconstructions of NGC~6505 (15~May) were made with slightly different
versions of this branch, but both
use the same S\'ersic source model and luminosity normalisation
(Sect.~\ref{sec:sersic}), so we expect the difference between them to have
a negligible effect on our results.

\paragraph{CosmoSim}\footnote{\url{https://github.com/CosmoAI-AES/CosmoSim}}
is an open-source
tool for visualising strong gravitational lenses and reconstructing lens
mass distributions from either simulated or observed images. One of the major features of CosmoSim is its ability to locally expand the lens potential around any point, in a so-called Roulette expansion. CosmoSim is built
around a three-layer architecture: A C++ simulation library performs the lens-equation
computations, including the derivation of Roulette amplitudes, critical
curves and the source light distribution; a Python wrapper, CosmoSimPy,
exposes this library to user-facing tools; and a set of Python scripts and
a graphical user interface (\texttt{CosmoGUI.py}) allow interactive or
batch generation of lensed images.
The simulator currently implements Singular Isothermal Sphere (SIS), 
Singular Isothermal Ellipsoid (SIE), and Point Mass lens models,
with spherical, elliptical and triangular source models.
It supports both ray-traced and Roulette-expansion simulation
modes; the latter reconstructs an image from a specified number of
Roulette terms. The source-light distribution is given by a S\'ersic
profile (Sect.~\ref{sec:sersic}), which lets the simulated source be
specified by an apparent magnitude and effective radius. CosmoSim also uses
a deferred, telescope-agnostic pixel-to-angular-scale calibration, in which pixel quantities are only
converted to angular units \emph{after} a reconstruction is complete, by
multiplying by the telescope's arcseconds-per-pixel scale $s_\text{arcsec}$
(e.g.\ 0.1\arcsec/pixel for Euclid VIS, 0.04\arcsec/pixel for HST/UVIS):
\begin{equation}
\theta_\text{E} = r_\text{pixel}\, s_\text{arcsec}, \qquad
\vec{\theta}_\text{arcsec} = \vec{x}_\text{sim}\, s_\text{arcsec}.
\label{eq:scale}
\end{equation}
This lets a single ML model, trained once on CosmoSim's dimensionless pixel
images, be applied unchanged to imaging from any survey, and, where the
redshifts of lens and source are known, lets the resulting angular Einstein
radius be converted to a physical Einstein radius and an enclosed mass
(Sect.~\ref{sec:results}).

\subsection{Training on synthetic data}
\label{sec:mltraining}

CosmoSim includes the \texttt{dataset} and \texttt{datagen} scripts,
which generate labelled parameter tables and the corresponding images in bulk. 
We used these tools to generate a large synthetic training set of SIE lens images
with the new S\'ersic source profile, spanning a broad range of Einstein
radii, ellipticities, orientations and source positions -- including
configurations both inside and outside the critical curves, so that the
training set contains doubly, triply and quadruply imaged systems in
proportions representative of the real observational sample. A
convolutional neural network (CNN) was trained exclusively on this synthetic data to
predict seven ray-trace parameters directly from an image: the lens
Einstein radius, ellipticity and orientation, and the source effective radius,
position angle and radius, and S\'ersic index. 

Parameter uncertainties are obtained with Monte Carlo dropout
\citep{Gal2016,Shirmohammadi2021}: each image is passed through the network
200 times with 20 per cent of the units randomly dropped, and the mean and
standard deviation of the predictions are taken as the estimate and its
uncertainty. Because the model never sees a real image during training, applying
it to observational data is a genuine out-of-distribution test of both the
ML model and of how realistic the underlying CosmoSim images are.

\paragraph{Training data} The training set comprises 30\,000 CosmoSim
images of SIE lenses with a spherical, S\'ersic-profile source, split
70/15/15 into training, validation and test sets, with parameter ranges
(Table~\ref{tbl:trainrange}) adapted from \citet{Hezaveh2017}. The source
brightness is set by the dimensionless total luminosity $L_\text{S\'ersic}$,
which is varied in the training data but not predicted by the ML model.

\begin{table}
\caption{Parameter ranges used to generate the synthetic training data.}
\label{tbl:trainrange}
\begin{tabular*}{\tblwidth}{@{}LLCL@{}}
\toprule
Symbol & Parameter & Range & Unit \\
\midrule
$\theta_\text{E}$      & Einstein radius     & 5--75      & pixels \\
$\epsilon$             & Ellipticity         & 0.5--1.0   & \\
$\phi_\text{L}$        & Lens orientation    & 0--180     & deg \\
$R_\text{Eff}$         & Source effective radius & 2--5   & pixels \\
$R_\text{S}$           & Source radius       & 1--10      & pixels \\
$\phi_\text{S}$        & Source angle        & 0--360     & deg \\
$n_\text{S\'ersic}$    & S\'ersic index      & 0.5--5.0   & \\
$L_\text{S\'ersic}$    & Total luminosity    & 10--20     & \\
\bottomrule
\end{tabular*}
\end{table}

\paragraph{Hyperparameters} The final model (a Wide ResNet50-2 backbone,
trained with a batch size of 96) was selected by a grid search over
hyperparameters including the learning rate and weight decay, choosing the
combination that minimised the validation loss. Training ran for at most
500 epochs with early stopping: training was halted once the validation loss
had not improved for 20 epochs (the patience), which limits overfitting and
saves computing time.

\subsection{Real lenses: Noise and decontamination}
\begin{figure}
    \subfloat[The lens fitted to system NGC 6505\label{fig:lenssub}]{%
        \includegraphics[width=0.31\linewidth]{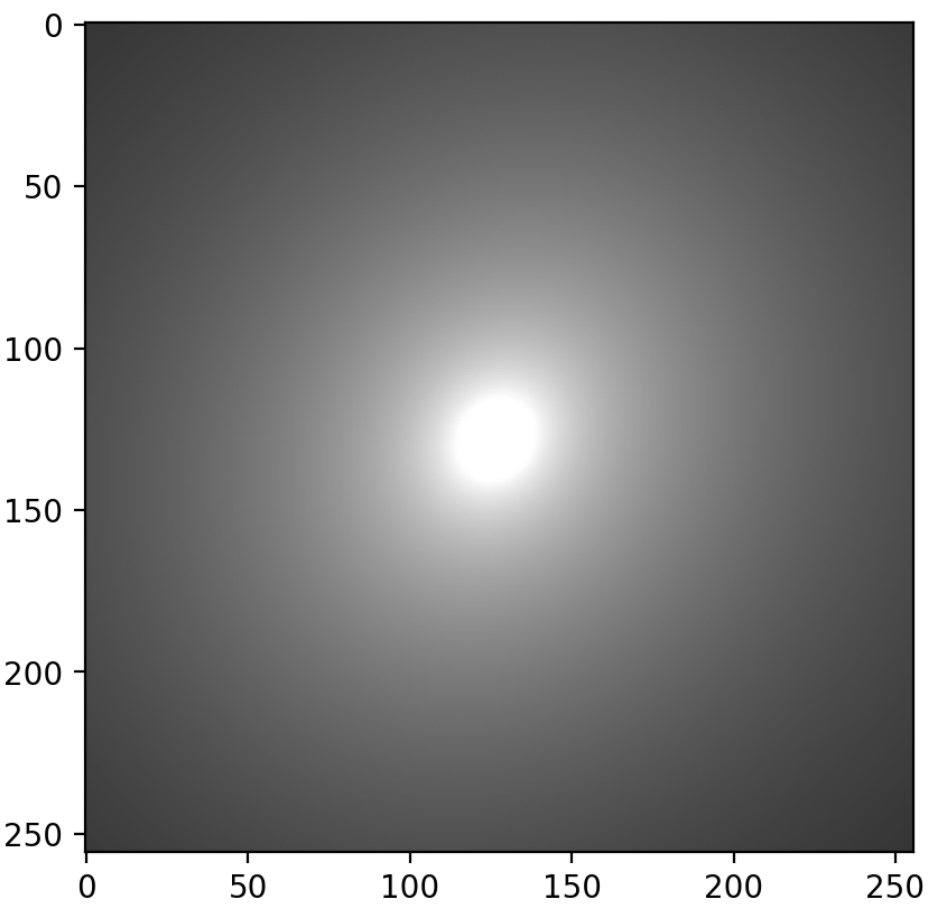}%
    }\hfill
    \subfloat[Mask of the remaining light from NGC 6505 after lens subtraction, with smoothing and a threshold value.\label{fig:softmask}]{%
        \includegraphics[width=0.31\linewidth]{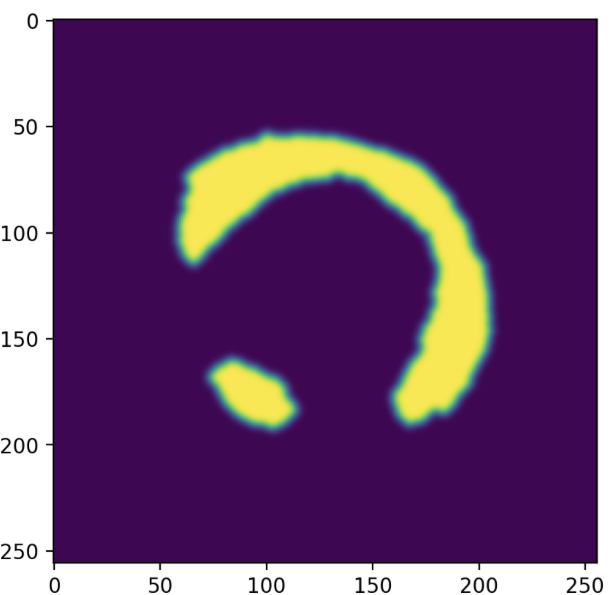}%
    }\hfill
    \subfloat[Final image after lens fitting and post processing \label{fig:final_cleanup}]{%
        \includegraphics[width=0.31\linewidth]{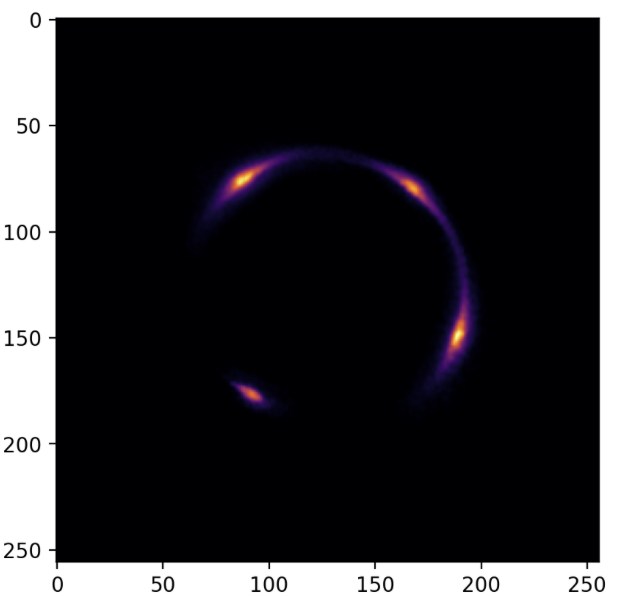}%
    }
  \caption{NGC~6505 lens subtraction: (a) the fitted lens light, (b) the
  mask of the remaining light after subtraction, with smoothing and a
  threshold value, and (c) the final image after lens fitting and
  post-processing.}
  \label{fig:ngc6505steps}
\end{figure}

\begin{figure}
  \centering
  \includegraphics[width=\linewidth]{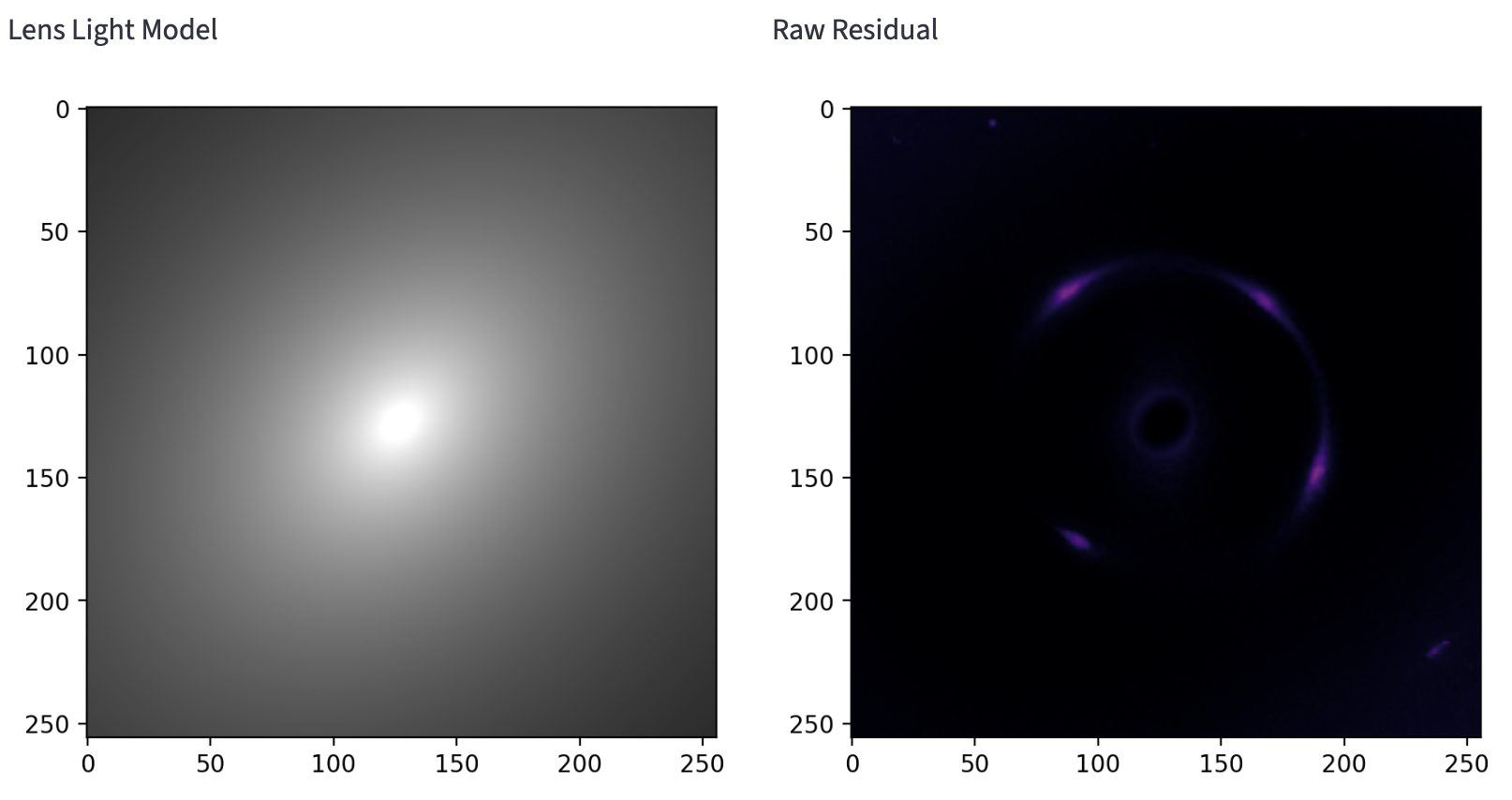}
  \caption{Example AstroPy fit: the modelled foreground lens light
  (left) and the raw residual obtained by subtracting it from the observed
  image (right), which reveals the lensed arcs. }
  \label{fig:lenslight}
\end{figure}

\begin{figure}
  \centering
  \includegraphics[width=0.8\linewidth]{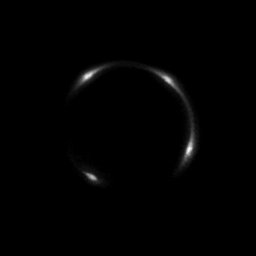}
  \caption{Decontaminated Euclid image of NGC~6505, with the foreground
  lens light subtracted using AstroPy, used as input to both the direct
  reconstruction and the machine-learning pipeline.}
  \label{fig:ngc6505clean}
\end{figure}

\begin{figure*}
  \centering
  \begin{tikzpicture}[
      node distance=5mm and 5mm,
      box/.style={draw, rounded corners=2pt, align=center, minimum height=8mm,
        text width=2.15cm, inner sep=2pt, fill=gray!7, font=\footnotesize},
      own/.style={box, fill=blue!12, draw=blue!50!black},
      arr/.style={-{Latex[length=2mm,width=1.5mm]}, thick}
    ]
    \node[box] (obs) {Real lens image\\(Euclid / AGEL)};
    \node[own, right=of obs] (clean) {AstroPy decontamination};
    \node[box, right=of clean] (cleaned) {Cleaned image};
    \node[box, right=of cleaned] (branch) {Literature params.\ \emph{or}\\ML prediction};
    \node[own, right=of branch] (cosmosim) {CosmoSim reconstruction};
    \node[box, right=of cosmosim] (recon) {Reconstructed image};

    \node[own, below=6mm of branch] (mltrain) {ML parameter estimator};
    \node[own, right=of mltrain] (synth) {CosmoSim training set (SIE + S\'ersic)};

    \draw[arr] (obs) -- (clean);
    \draw[arr] (clean) -- (cleaned);
    \draw[arr] (cleaned) -- (branch);
    \draw[arr] (branch) -- (cosmosim);
    \draw[arr] (cosmosim) -- (recon);
    \draw[arr] (cleaned.south) |- (mltrain.west);
    \draw[arr] (mltrain) -- (branch);
    \draw[arr] (synth) -- (mltrain);
    \draw[arr, dashed, draw=orange!70!black] (recon.north) -- ++(0,4mm) -| (cleaned.north)
      node[pos=0.25, above, font=\scriptsize, text=orange!60!black]
      {Comparison metrics: SSIM, IoU, Hausdorff, Chamfer};
  \end{tikzpicture}
  \caption{The full CosmoSim reconstruction pipeline. A real, decontaminated
  survey image is reconstructed in CosmoSim either from literature lens
  parameters or from parameters predicted by an ML model; the ML model is
  trained only on CosmoSim's own synthetic images (bottom row). The
  reconstruction is compared with the cleaned image using several
  morphology-sensitive metrics. Blue boxes mark the tools and
  data sets developed by the authors.}
  \label{fig:flow}
\end{figure*}
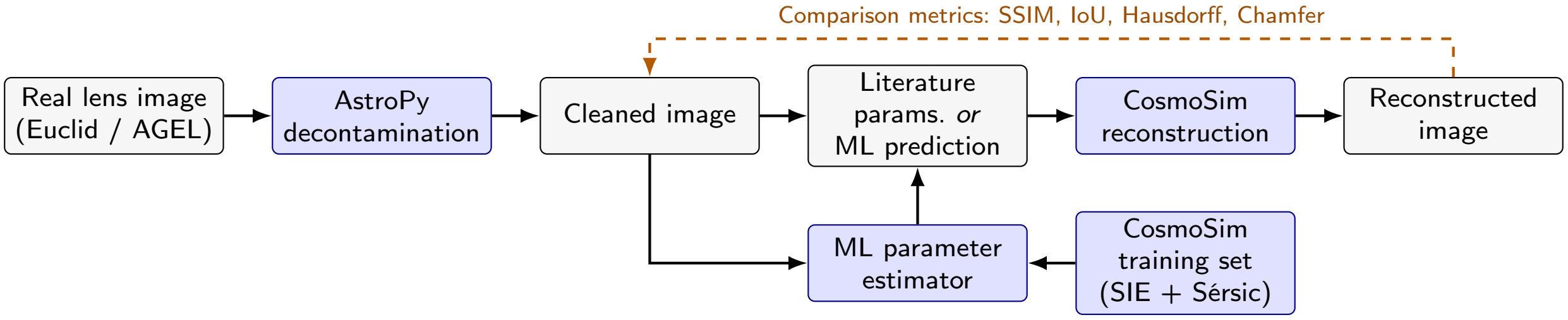

To test CosmoSim and the ML estimator of Sect.~\ref{sec:mltraining} against
real data, we chose the newly discovered Euclid Einstein Cross NGC~6505
\citep{ORiordan2025} alongside 20 further images of 18 systems: 12 from the
AGEL survey~\citep{Tran2022} and 8 images of 6 systems from the Euclid
Strong Lensing Discovery Engine \citep{EuclidSLDE2025}. For two of the
Euclid systems the lens light was removed in two versions, retaining one or
two images of the source, respectively; both versions are kept as separate
entries (Euclid-3/Euclid-5 and Euclid-4/Euclid-6 in
Table~\ref{tab:targetnames}). Before a real image
can be compared with, or fed into, CosmoSim, the foreground lens galaxy's
light must be removed so that only the lensed source remains. Working
with lens-light-subtracted images follows \citet{Hezaveh2017}, who applied
convolutional networks to real lenses after removing the lens light; unlike
them, however, we do not add observational noise to the synthetic training
images, but instead process the real images so that they resemble the
noise-free CosmoSim output, as described below. We developed a graphical tool for removing lens-galaxy light: it fits the lens with one or more Sérsic components using AstroPy \citep{Astropy2022}, subtracts the model, and then isolates the lensed arcs by applying Gaussian smoothing and a threshold mask. Figure~\ref{fig:lenslight} shows an example of fitted lens-light
model and the resulting raw residual immediately after subtraction; visual
inspection of this residual is used to verify the quality of the fit before
the lensed arcs are isolated. The raw residual is then cleaned up by
masking out unrelated objects and enforcing a minimum object size and flux
threshold, and the resulting mask is smoothed to soften its edges before
being applied to the residual, producing a
final image with an arc morphology closer to CosmoSim's own simulated
output. For NGC~6505, the best lens subtraction was obtained with the
three S\'ersic components listed in Table~\ref{tbl:ngc6505fit}.
Fig.~\ref{fig:ngc6505steps} shows the resulting fitted lens light, the mask
of the remaining light after subtraction, and the final post-processed
image, while Fig.~\ref{fig:ngc6505clean} shows the same final image
rescaled for use as input to the reconstruction and ML pipelines.
Figure~\ref{fig:flow} summarises the full pipeline, from a real survey
image to a reconstructed image and its comparison metrics.

\begin{table}
\caption{S\'ersic components used to fit and subtract the lens light of
NGC~6505.}
\label{tbl:ngc6505fit}
\begin{tabular*}{\tblwidth}{@{}LCCC@{}}
\toprule
Component & Amplitude & $R_\text{Eff}$ (pixels) & $n_\text{S\'ersic}$ \\
\midrule
1 & 1.0 & 40  & 1.00 \\
2 & 0.3 & 70  & 1.35 \\
3 & 0.1 & 125 & 4.00 \\
\bottomrule
\end{tabular*}
\end{table}

\paragraph{Comparison metrics} We compare each reconstruction with the
cleaned image using four complementary metrics. The structural similarity
index (SSIM; \citealt{Wang2004}) compares local brightness, contrast and
structure in a sliding window and averages the result, with 1 for identical
images. We evaluate it both over the full image (\emph{global} SSIM) and
only over the non-zero pixels, i.e.\ the lensed arcs (\emph{masked} or
\emph{local} SSIM). The remaining three metrics compare the shapes of the
arcs, represented by binary masks obtained by thresholding the two images.
The intersection-over-union (IoU) is the area where the two masks overlap
divided by the area covered by either, with 1 for perfect overlap. The
Hausdorff distance is the largest distance from a point in one mask to the
nearest point in the other, and so measures the worst mismatch, while the
Chamfer distance averages the distance to the nearest point in the other
mask and is less sensitive to single outliers; both are given in pixels and
vanish for identical masks.

\section{Results and analysis}
\label{sec:results}
In the following we present the results on our main target of investigation, the Einstein Cross NGC~6505, alongside 20 additional images on which our reconstruction pipeline was applied.

\subsection{NGC~6505}
\label{sec:ngc6505results}

\paragraph{Reconstruction from literature parameters}

\begin{figure}
  \centering
      \subfloat[\label{fig:lensdirect}]{%
        \includegraphics[width=0.47\linewidth]{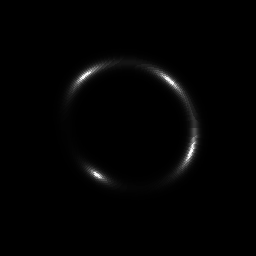}%
    }\hfill
    \subfloat[\label{fig:lensdirectdiff}]{%
        \includegraphics[width=0.47\linewidth]{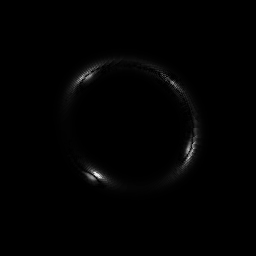}%
    }
  \caption{CosmoSim reconstruction of NGC~6505 using the SIE lens
  parameters derived from the EPL fit of \citet{ORiordan2025} (left) and
  the pixel-by-pixel difference with respect to the cleaned Euclid image of
  Fig.~\ref{fig:ngc6505clean} (right).}
  \label{fig:ngc6505recon}
\end{figure}

Feeding the Elliptical Power Law lens parameters (with slope $\gamma \approx 2$) reported by
\citet{ORiordan2025} directly into CosmoSim as a SIE lens
(Table~\ref{tbl:ngc6505lit}; the source position and light-profile
parameters were not published and were tuned manually) reproduces the
observed four-fold image structure with high visual fidelity
(Fig.~\ref{fig:ngc6505recon}): a correctly positioned quadruple Einstein
Cross, with only slight positional offsets relative to the decontaminated
Euclid data (Fig.~\ref{fig:ngc6505clean}). Quantitatively
(Table~\ref{tbl:ngc6505}, row ``Literature''), this reconstruction scores a
masked SSIM of 0.54, an IoU of 0.45, a Chamfer distance of 5.5~pixels and a
Hausdorff distance of 11.5~pixels against the cleaned image. We attribute the residual mismatch to two simplifications. First, the
source position and light profile were not published and had to be tuned by
hand. Second, CosmoSim's SIE is only an approximation to the discovery
team's model: their Elliptical Power Law (EPL) reduces to the SIE for a
power-law slope $\gamma=2$, and although their slope ($\gamma=2.03$) is
close to this, their preferred model also includes external shear and an
$m=4$ multipole \citep{ORiordan2025}, which the SIE lacks. Such
limitations of a fixed parametric form illustrate the potential value of
parameter-free modelling.

\begin{table}
\caption{Literature SIE lens and source parameters used to reconstruct
NGC~6505 \citep[their preferred model, Table~2]{ORiordan2025}. $R_\text{S}$,
$\phi_\text{S}$ and $R_\text{Eff}$ were not reported and were tuned manually.
The pixel scale of the simulated image is 0.0390625\arcsec/pixel.}
\label{tbl:ngc6505lit}
\begin{tabular*}{\tblwidth}{@{}LLCL@{}}
\toprule
Symbol & Parameter & Value & Unit \\
\midrule
$\theta_\text{E}$  & Einstein radius         & 64    & pixels \\
$\phi_\text{L}$    & Lens orientation        & 143.3 & deg \\
$\epsilon$         & Ellipticity             & 0.702  & \\
$R_\text{S}$       & Source radius           & 3.8  & pixels \\
$\phi_\text{S}$    & Source angle            & 77.5 & deg \\
$R_\text{Eff}$     & Source effective radius & 3.47 & pixels \\
\bottomrule
\end{tabular*}
\end{table}

\begin{figure}
  \centering
      \subfloat[\label{fig:ngcml}]{%
        \includegraphics[width=0.47\linewidth]{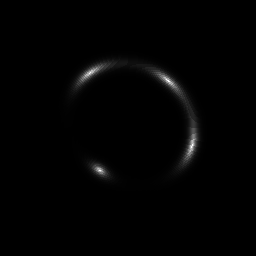}%
    }\hfill
    \subfloat[\label{fig:ngcmldiff}]{%
        \includegraphics[width=0.47\linewidth]{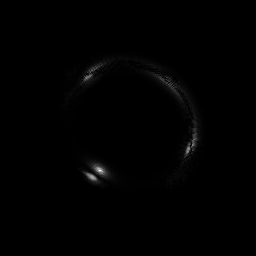}%
    }
  \caption{Cleaned Euclid image of NGC~6505 (left) and CosmoSim
  reconstruction (right) using lens and source parameters predicted by the
  machine-learning model, with no manual adjustment and no access to the
  literature parameters of \citet{ORiordan2025}.}
  \label{fig:ngc6505ml}
\end{figure}

\paragraph{Reconstruction from ML-predicted parameters}
Without using any literature values, the ML model of
Sect.~\ref{sec:mltraining} estimated the lens and source parameters
directly from the cleaned image of Fig.~\ref{fig:ngc6505clean}
(Table~\ref{tbl:ngc6505ml}): Einstein radius $61.7\pm5.5$~pixels,
ellipticity $0.63\pm0.04$ and lens orientation
$142.7\degr\pm14.7\degr$. Feeding these predictions back into CosmoSim
again reproduces the ring-like structure with its four brightness knots
(Fig.~\ref{fig:ngc6505ml}): the reconstruction correctly recovers the
spatial position of three of the four images, with the fourth showing an
inward radial offset and un-subtracted flux, consistent with the residual
ellipticity error of the prediction. This reconstruction -- built without
any access to the discovery team's measurements and with no manual tuning --
performs at least as well as the one built directly from their published
parameters (Table~\ref{tbl:ngc6505}, row ``ML-predicted''): it scores better
on masked SSIM (0.65 versus 0.54), IoU (0.55 versus 0.45) and Chamfer
distance (4.1 versus 5.5~pixels), while the Hausdorff distances are
essentially equal (11.7 versus 11.5~pixels). A masked SSIM well below unity
indicates that pixel-level flux agreement within the arcs is still limited,
even though their overall shape and position are well recovered.

\begin{table}
\caption{Lens and source parameters for NGC~6505 predicted by the ML model
with Monte Carlo dropout uncertainties over 200 forward passes
(Sect.~\ref{sec:mltraining}).}
\label{tbl:ngc6505ml}
\begin{tabular*}{\tblwidth}{@{}LLCL@{}}
\toprule
Symbol & Parameter & Value & Unit \\
\midrule
$R_\text{S}$       & Source radius    & $6.02\pm0.65$    & pixels \\
$\phi_\text{S}$    & Source angle     & $69.54\pm35.71$  & deg \\
$\theta_\text{E}$  & Einstein radius  & $61.66\pm5.49$   & pixels \\
$\epsilon$         & Ellipticity      & $0.63\pm0.04$    & \\
$\phi_\text{L}$    & Lens orientation & $142.72\pm14.71$ & deg \\
$R_\text{Eff}$     & Source effective radius & $3.81\pm0.25$ & pixels \\
$n_\text{S\'ersic}$ & S\'ersic index & $0.90\pm0.37$ & \\
\bottomrule
\end{tabular*}
\end{table}

\paragraph{Summary of comparison metrics}
Table~\ref{tbl:ngc6505} collects the comparison metrics for the two
reconstructions side by side. The Chamfer distance is insensitive to the
exact pixel-flux values and instead compares the geometric boundaries of the
lensed arcs.

\begin{table}
\caption{Image-comparison metrics for the two CosmoSim reconstructions of
NGC~6505, evaluated against the decontaminated Euclid image. SSIM is
evaluated over the lensed arcs only (masked SSIM); the Chamfer and Hausdorff
distances are in pixels. Higher SSIM and IoU, and smaller distances,
indicate better agreement.}
\label{tbl:ngc6505}
\begin{tabular*}{\tblwidth}{@{}LCCCC@{}}
\toprule
Reconstruction & Chamf. & SSIM (local) & IoU & Hausd.\ \\
\midrule
Literature        & 5.464 & 0.542 & 0.449 & 11.5 \\
ML-predicted       & 4.144 & 0.654 & 0.550 & 11.7 \\
\bottomrule
\end{tabular*}
\end{table}

Overall, both reconstruction routes -- literature parameters and
ML-predicted parameters -- recover the correct qualitative topology of the
Einstein Cross, and the ML-predicted reconstruction performs at least as
well: it has higher masked SSIM and IoU and a smaller Chamfer distance,
whereas the literature reconstruction has a marginally smaller Hausdorff
distance (Table~\ref{tbl:ngc6505}). Since the ML model had no access whatsoever to the discovery team's
own measurements, this is a positive validation of the full CosmoSim/ML
pipeline, and of CosmoSim's synthetic training images as a usable proxy for
real Euclid data. The residual errors in both cases point to the same root
cause: parametric reconstructions will never be perfect, precisely because they are built from parameter models that do not fully correspond to the real world.

\subsection{Ensemble of additional strong lenses}
\label{sec:ensembleresults}
\begin{figure}
  \centering
  \includegraphics[width=\linewidth]{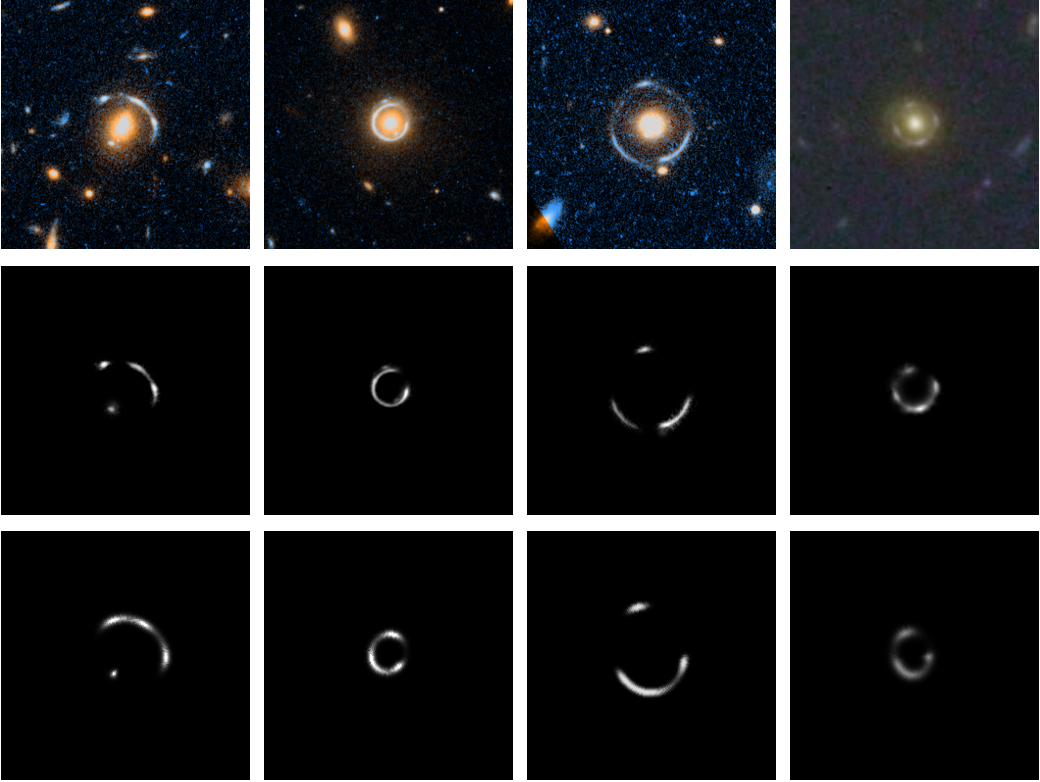}
  \caption{Four qualitatively successful reconstructions from the
  observational ensemble. Top row: original survey images. Middle row:
  decontaminated versions. Bottom row: final CosmoSim reconstructions.}
  \label{fig:success}
\end{figure}

\begin{figure}
  \centering
  \includegraphics[width=\linewidth]{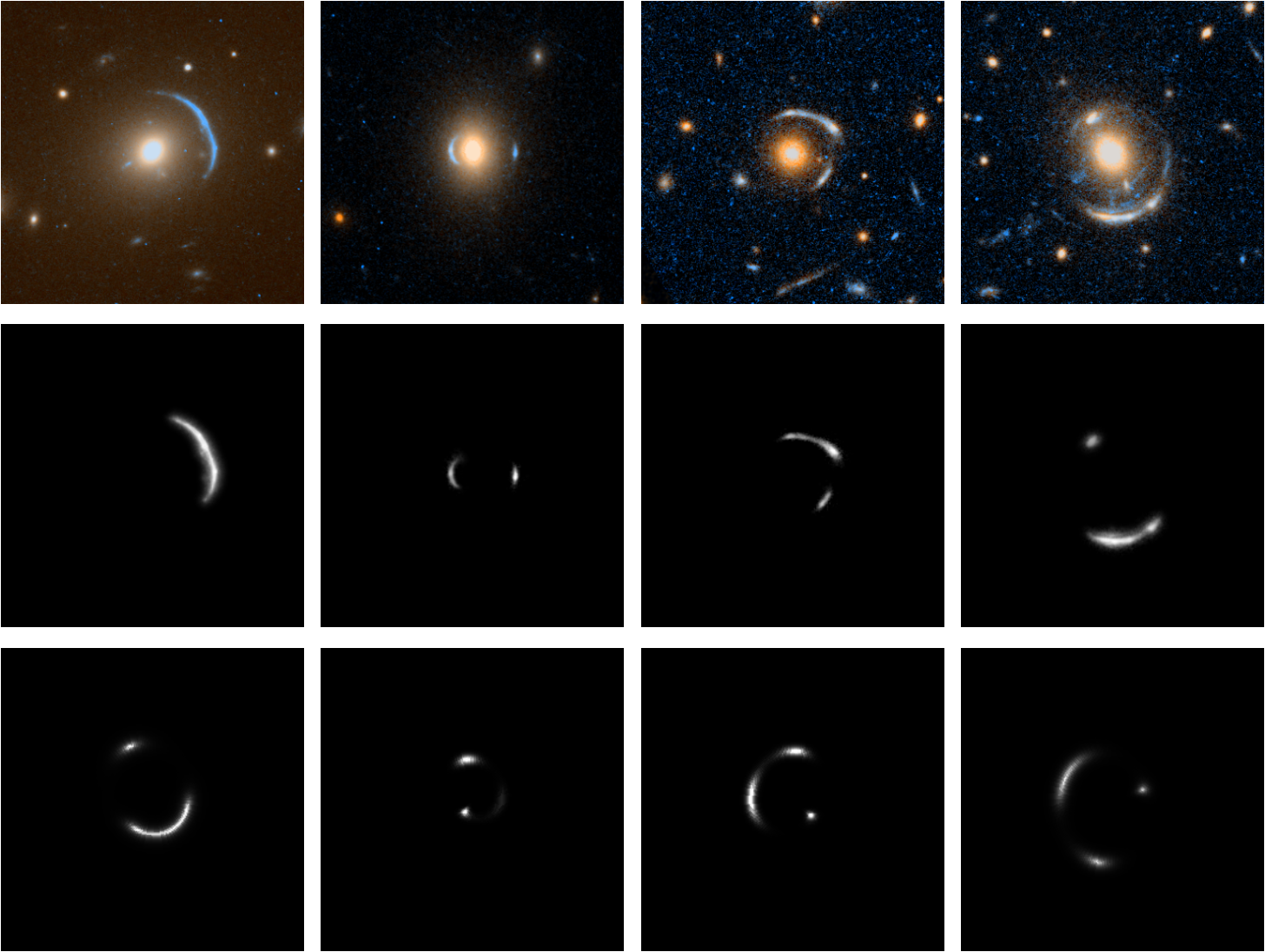}
  \caption{Four qualitatively unsuccessful reconstructions, where the model
  fails to capture the core topology of the lensed arcs. Rows as in
  Fig.~\ref{fig:success}.}
  \label{fig:fail}
\end{figure}

Beyond NGC~6505, a further 20 images of 18 strong-lens systems from the
AGEL survey and the Euclid Strong Lensing Discovery Engine were
lens-subtracted, decontaminated and reconstructed with the same pipeline,
for an ensemble of 21 images in total. Figure~\ref{fig:success} shows four qualitatively
successful reconstructions, clearly recognisable to the human eye even
where the size or position of the arcs shows a small spatial offset.
Figure~\ref{fig:fail} shows four qualitatively unsuccessful cases, where the
model fails to capture the core topology of the lensed arcs -- these are
predominantly the systems with the most complex arc/ring morphology or the
lowest signal-to-noise ratio after decontamination.

Averaging the SSIM score across the full ensemble reveals a large gap
between the global and masked (arc-only) values: a global SSIM of 0.95
compared with a masked SSIM of only 0.24. As for NGC~6505 individually,
this shows that the global score is dominated by the shared, mostly empty
background, and that the masked score is a much stricter, and more
representative, measure of how well the reconstructed arcs actually agree
with the observed ones; the same caveat applies to the IoU score, which is
similarly sensitive to small positional offsets between otherwise
well-matched arcs. Bypassing this spatial sensitivity with the Chamfer
distance -- which compares the geometric boundaries of the lensed arcs
directly -- generally shows a high degree of morphological similarity
between observation and reconstruction, even in cases penalised harshly by
SSIM/IoU, though it also exposes cases where CosmoSim merges several
separate, distinct lensed features (as resolved in the real image) into a
single continuous arc.

\begin{table}
\caption{Ensemble-averaged image-comparison metrics for the ML model.}
\label{tbl:model1v2}
\begin{tabular*}{\tblwidth}{@{}LC@{}}
\toprule
Metric & ML model \\
\midrule
Local SSIM         & 0.24  \\
Chamfer distance     & 16.54 \\
Hausdorff distance    & 23.16 \\
IoU                   & 0.19 \\
\bottomrule
\end{tabular*}
\end{table}

\paragraph{Physical parameters: Euclid systems}
Table~\ref{tbl:euclid} compares the predicted Einstein radius, ellipticity
and lens orientation for the Euclid-discovered systems with the catalogued
PyAutoLens models (SIE plus external shear) of the Euclid Q1 Strong Lensing
Discovery Engine \citep{EuclidSLDE2025,SLDEdata2025}. Catalogued models are
available for seven of the eight Euclid entries (Euclid-8 has none), and
Euclid-3/Euclid-5 and Euclid-4/Euclid-6 are two versions of the same
systems, so the table contains five distinct systems. The catalogued values
are themselves model estimates with purely statistical uncertainties, not
ground truth, so the comparison tests the consistency of two independent
modelling approaches. The predicted Einstein radii agree with the catalogued
ones within the $1\sigma$ prediction uncertainty for all systems (mean
relative difference 0.08). The predicted axis ratios, by contrast, differ
systematically from the catalogued values: they are lower (0.44--0.54 versus
0.61--0.88, by more than $5\sigma$ for four of the seven entries). Since the
catalogued values are themselves model estimates, this comparison alone
cannot establish which is correct. Several observations nevertheless point
to a limitation of the ML model rather than of the catalogue. (i) Four of the predicted axis ratios lie below 0.5, the
lower edge of the training range (Table~\ref{tbl:trainrange}), suggesting
that the real images differ from the synthetic training images. (ii) The
catalogue models are fitted at pixel level to the original images including
the PSF, whereas the ML model sees images that have been lens-subtracted,
smoothed and masked, processing that may alter the shape of the arcs.
(iii) The catalogued axis ratios are typical of massive elliptical lens
galaxies, whereas all seven predictions are unusually flattened. Conversely,
the catalogue models include external shear, which is partly degenerate
with ellipticity \citep{Keeton1997}. A model without shear, such as CosmoSim's
SIE, must absorb an external shear $\gamma$ aligned with the lens into a
flatter mass distribution, lowering the axis ratio by an amount of order a
few times $\gamma$.
The catalogued shears (0.02--0.12) can account for most of the difference
for Euclid-4/Euclid-6, which has the largest shear ($\gamma\approx0.12$),
nearly aligned with the lens, and for part of it for Euclid-2 and Euclid-7,
but not for Euclid-1, where the shear is small, nor for Euclid-3/Euclid-5,
where it is nearly perpendicular to the lens. External shear thus explains
part, but not all, of the systematic offset.

The comparison of lens orientations is less conclusive. CosmoSim and the
catalogue define the angle relative to different axes of the ellipse (minor
versus major axis; Sect.~\ref{sec:sie}), which we have corrected for, but
we have not been able to establish whether the two codes also orient the
image axes identically, in particular whether the image $y$-axis points up
or down. With the conversion adopted in Table~\ref{tbl:euclid}, the
orientations differ by $53^\circ$ on average, comparable to the
$45^\circ$ expected for random orientations; with a reversed
$y$-axis it would be $64^\circ$. Only if, in addition, CosmoSim's
$\phi_\text{L}$ referred to the major rather than the minor axis would the
mean difference drop to $26^\circ$, with four of the seven entries agreeing
to within $10^\circ$. The orientation is moreover poorly defined
for the nearly round systems with $q\gtrsim0.8$. We therefore draw no
conclusion about how well the ML model recovers the lens orientation.

\begin{table*}
\caption{Euclid ensemble: predicted vs.\ catalogued Einstein radius
$\theta_\text{E}$ (arcsec), ellipticity $\epsilon$ and lens orientation
$\phi_\text{L}$ (deg). Catalogued values are the PyAutoLens SIE models of the
Euclid Q1 Strong Lensing Discovery Engine \citep{EuclidSLDE2025,SLDEdata2025}:
$\theta_\text{E}$ is their effective Einstein radius and $\epsilon$ their
axis ratio $q$. Their position angle refers to the major axis, whereas
CosmoSim's $\phi_\text{L}$ refers to the minor axis (both counter-clockwise
from the image $x$-axis, Sect.~\ref{sec:sie}); the catalogued angles have
therefore been converted as $\phi_\text{L}=\phi_\text{cat}+90^\circ$
(mod $180^\circ$). RD is the relative difference $|\text{pred}-\text{cat}|/\text{cat}$;
for the orientation we give instead the absolute difference
$|\Delta\phi_\text{L}|$ (mod $180^\circ$). Whether CosmoSim and PyAutoLens orient the image axes identically has not
been established, so the orientation comparison should be treated with
caution (see text).
Euclid-3/Euclid-5 and Euclid-4/Euclid-6 are the same systems with
different lens-light subtraction; Euclid-8 has no catalogued model and is
omitted. The lensing systems have been given an arbitrary order.}
\label{tbl:euclid}
\begin{tabular*}{\tblwidth}{@{}LCCCCCCCCC@{}}
\toprule
Object & $\theta_\text{E}$ pred & $\theta_\text{E}$ lit. & RD & $\epsilon$ pred & $\epsilon$ lit. & RD & $\phi_\text{L}$ pred & $\phi_\text{L}$ lit. & $|\Delta\phi_\text{L}|$ \\
\midrule
Euclid-1 & $0.80 \pm 0.239$ & $0.80 \pm 4.6\times10^{-3}$ & 0.00 & $0.46 \pm 0.053$ & $0.78 \pm 0.019$ & 0.41 & $99.79  \pm 13.629$ & $20.18 \pm 3.067$ & 79.6 \\
Euclid-2 & $1.48 \pm 0.229$ & $1.45 \pm 9.4\times10^{-3}$ & 0.02 & $0.49 \pm 0.047$ & $0.82 \pm 0.036$ & 0.40 & $72.55  \pm 15.088$ & $27.35 \pm 6.578$ & 45.2 \\
Euclid-3 & $1.04 \pm 0.205$ & $1.20 \pm 21.9\times10^{-3}$& 0.13 & $0.54 \pm 0.046$ & $0.61 \pm 0.041$ & 0.11 & $108.48 \pm 14.203$ & $159.20 \pm 3.589$ & 50.7 \\
Euclid-4 & $0.87 \pm 0.265$ & $1.07 \pm 6.2\times10^{-3}$ & 0.19 & $0.50 \pm 0.045$ & $0.88 \pm 0.015$ & 0.44 & $72.03  \pm 14.006$ & $14.90 \pm 3.310$ & 57.1 \\
Euclid-5 & $1.19 \pm 0.262$ & $1.20 \pm 21.9\times10^{-3}$& 0.01 & $0.44 \pm 0.059$ & $0.61 \pm 0.041$ & 0.28 & $117.93 \pm 15.470$ & $159.20 \pm 3.589$ & 41.3 \\
Euclid-6 & $1.19 \pm 0.313$ & $1.07 \pm 6.2\times10^{-3}$ & 0.11 & $0.45 \pm 0.061$ & $0.88 \pm 0.015$ & 0.49 & $115.26 \pm 16.049$ & $14.90 \pm 3.310$ & 79.6 \\
Euclid-7 & $1.21 \pm 0.240$ & $1.39 \pm 8.9\times10^{-3}$ & 0.13 & $0.53 \pm 0.054$ & $0.64 \pm 0.021$ & 0.16 & $101.70 \pm 15.680$ & $81.59 \pm 0.637$ & 20.1 \\
\bottomrule
\end{tabular*}
\end{table*}

\paragraph{Physical parameters: AGEL systems}
Table~\ref{tbl:agel} compares the predicted Einstein radius and enclosed
mass $M(<\theta_\text{E})$ for the 12 AGEL systems with the values of
\citet{Gottemoller2026}. These catalogued Einstein radii come from simple
ring or image-pair fits to ground-based imaging and carry uncertainties of
0.7--1.2\arcsec\ (30--50 per cent), so they are considerably less precise
than the Euclid values. The mean relative difference in $\theta_\text{E}$ is
0.19, but all predictions agree with the catalogued values within the
combined uncertainties (the largest difference, AGEL-1, is $1.6\sigma$).
Since the enclosed mass is computed from $\theta_\text{E}$ using the
catalogued lens and source redshifts, $M(<\theta_\text{E})\propto\theta_\text{E}^2$,
differences in $\theta_\text{E}$ carry over directly to the mass: the
predicted masses differ from the catalogued values by 0.01--0.56~dex
(median 0.12~dex), again within the combined uncertainties (at most
$1.8\sigma$). The AGEL lenses are predominantly group- and cluster-embedded
\citep{Gottemoller2026}, so neither CosmoSim's single-galaxy SIE nor a
simple ring fit need describe them fully; the larger scatter relative to
the Euclid sample therefore reflects both the less precise catalogue values
and the more complex lens environments.

\begin{table*}
\caption{AGEL ensemble: predicted vs.\ catalogued Einstein radius $\theta_\text{E}$
(arcsec) and enclosed mass $M(<\theta_\text{E})$ (log$_{10}\,M_\odot$).
Catalogued values are from \citet{Gottemoller2026}. For $\theta_\text{E}$ we give
the relative difference (RD, as in Table~\ref{tbl:euclid}); for the mass, the absolute difference
$|\Delta\log_{10}M|$ in dex. The lensing systems have been arbitrarily ordered.}
\label{tbl:agel}
\begin{tabular*}{\tblwidth}{@{}LCCCCCC@{}}
\toprule
Object & $\theta_\text{E}$ pred & $\theta_\text{E}$ lit. & RD & $M$ pred & $M$ lit. & $|\Delta\log_{10}M|$ \\
\midrule
AGEL-1  & $1.96 \pm 0.526$ & $3.9 \pm 1.1$ & 0.50 & $12.04 \pm 0.23$ & $12.6 \pm 0.2$ & 0.56 \\
AGEL-2  & $2.35 \pm 0.379$ & $2.2 \pm 1.0$ & 0.07 & $12.22 \pm 0.14$ & $12.2 \pm 0.3$ & 0.02 \\
AGEL-3  & $3.35 \pm 0.425$ & $3.1 \pm 1.2$ & 0.08 & $12.64 \pm 0.11$ & $12.6 \pm 0.3$ & 0.04 \\
AGEL-4  & $1.73 \pm 0.435$ & $1.7 \pm 0.7$ & 0.02 & $12.19 \pm 0.22$ & $12.2 \pm 0.3$ & 0.01 \\
AGEL-5  & $2.58 \pm 0.449$ & $2.4 \pm 1.1$ & 0.08 & $12.37 \pm 0.15$ & $12.3 \pm 0.3$ & 0.07 \\
AGEL-6  & $1.33 \pm 0.587$ & $2.1 \pm 0.9$ & 0.37 & $11.42 \pm 0.38$ & $11.8 \pm 0.3$ & 0.38 \\
AGEL-7  & $1.90 \pm 0.549$ & $2.4 \pm 0.8$ & 0.21 & $11.79 \pm 0.25$ & $12.0 \pm 0.3$ & 0.21 \\
AGEL-8  & $2.64 \pm 0.448$ & $3.5 \pm 1.0$ & 0.25 & $12.39 \pm 0.15$ & $12.6 \pm 0.2$ & 0.21 \\
AGEL-9  & $2.94 \pm 0.505$ & $3.5 \pm 1.0$ & 0.16 & $12.08 \pm 0.15$ & $12.2 \pm 0.2$ & 0.12 \\
AGEL-10 & $3.44 \pm 0.554$ & $3.2 \pm 1.2$ & 0.07 & $12.51 \pm 0.14$ & $12.4 \pm 0.3$ & 0.11 \\
AGEL-11 & $1.49 \pm 0.405$ & $1.7 \pm 0.8$ & 0.12 & $11.71 \pm 0.24$ & $11.8 \pm 0.3$ & 0.09 \\
AGEL-12 & $2.80 \pm 0.483$ & $2.1 \pm 0.9$ & 0.33 & $12.41 \pm 0.15$ & $12.1 \pm 0.3$ & 0.31 \\
\bottomrule
\end{tabular*}
\end{table*}

The catalogue identifiers of the 12 AGEL systems and the 8 Euclid entries (6 distinct systems) are listed in Table~\ref{tab:targetnames}.

\section{Conclusion}
\label{sec:conclusion}
In this paper we have presented the first application of the CosmoSim
toolbox to real data, using a machine-learning model trained on synthetic
CosmoSim images to predict lens and source parameters directly from
observational images. We have validated the full pipeline against 21 images
of 19 real strong lenses from Euclid and the AGEL survey, after subtracting the foreground lens light with a custom decontamination tool built on AstroPy. As a case study we have
reconstructed the Einstein Cross NGC~6505, discovered by Euclid, both from the Elliptical Power Law lens parameters reported by the discovery team and from parameters predicted directly by our ML model. Both reconstructions recover the observed four-fold image structure against the decontaminated Euclid image.

The ML-predicted reconstruction, built with no access to the discovery team's own measurements, performs at least as well as the one built directly from their published lens model: it scores better on masked structural similarity, intersection-over-union (0.55 versus 0.45) and Chamfer distance (4.1 versus 5.5 pixels), while the Hausdorff distances are essentially equal (11.7 versus 11.5 pixels). Across the wider ensemble, the predicted Einstein radius is consistent with catalogued values within the quoted uncertainties for all systems (mean relative difference 0.08 for Euclid systems and 0.19 for the group- and cluster-embedded AGEL lenses), and the corresponding enclosed masses agree to within a median of 0.12~dex. The predicted lens ellipticity, by contrast, differs systematically from the catalogued values; the evidence points to a limitation of the ML model on real images, and testing this directly is a target for future work. The comparison of lens orientations is inconclusive, since we have not been able to establish that CosmoSim and the catalogue use identical angle conventions. These results establish CosmoSim, for the first time, as a tool that can be applied directly to observational data rather than only to synthetic images.

\printcredits

\bibliographystyle{cas-model2-names}
\bibliography{refs}

\appendix

\section{Decontaminated and reconstructed image pairs}
\label{app:ensemble}

\begin{figure*}
  \centering
  \includegraphics[width=\textwidth]{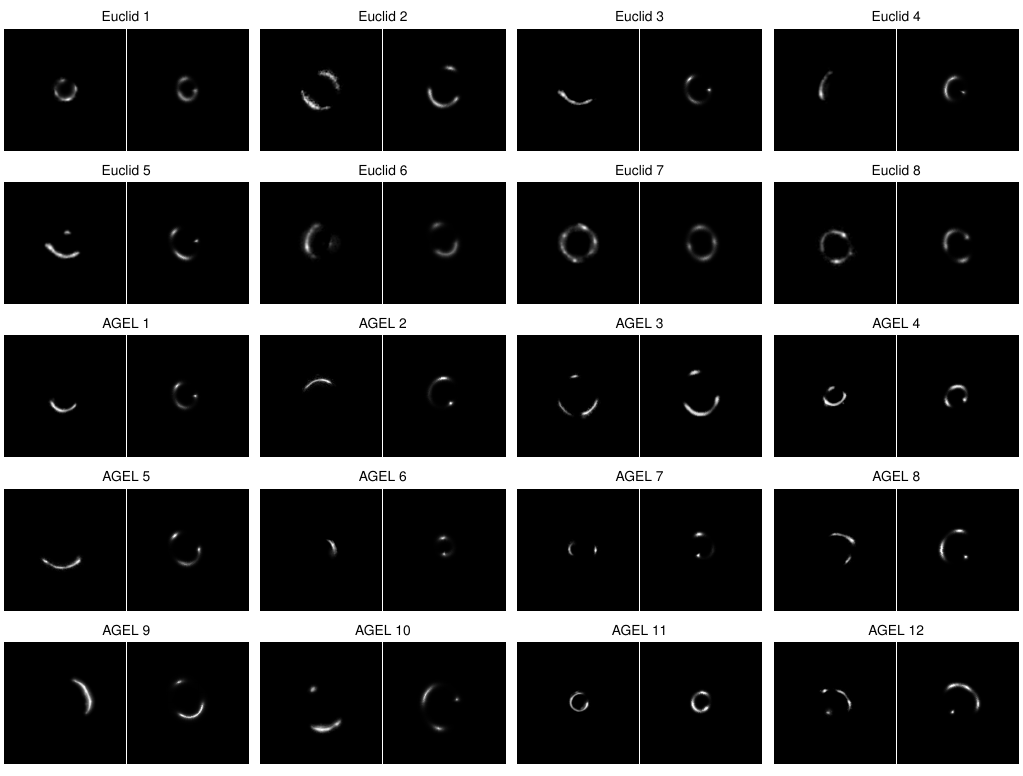}
  \caption{Decontaminated (left in each pair) and CosmoSim-reconstructed
  (right in each pair) images for the observational ensemble.}
  \label{fig:ensemblepairs}
\end{figure*}

Figure~\ref{fig:ensemblepairs} shows, for every system in the
observational ensemble, the decontaminated survey image next to its
CosmoSim reconstruction from the ML-predicted parameters. Labels follow
Table~\ref{tab:targetnames}.

\begin{table}[htbp]
\centering
\begin{tabular}{c c}
\toprule
\toprule
\textbf{Label} &
\textbf{Target ids} \\
\midrule
Euclid 1 & 102158584\_2687900321650859511 \\ [1.1ex]
Euclid 2 & 102158887\_2646461630654930628 \\ [1.1ex]
Euclid 3 & 102023478\_NEG628457640459296313 \\ [1.1ex]
Euclid 4 & 102158275\_2709777393645726835 \\ [1.1ex]
Euclid 5 & 102023478\_NEG628457640459296313 \\ [1.1ex]
Euclid 6 & 102158275\_2709777393645726835 \\ [1.1ex]
Euclid 7 & 102159773\_2677079122669151496 \\ [1.1ex]
Euclid 8 & 102021017\_NEG634146279486865014 \\ [1.1ex]
AGEL 1 & DESJ0141-1713-1 \\ [1.1ex]
AGEL 2 & DESJ0142-1831 \\ [1.1ex]
AGEL 3 & DESJ2125-6504 \\ [1.1ex]
AGEL 4 & DESJ0102+0158 \\ [1.1ex]
AGEL 5 & DESJ2219-4348 \\ [1.1ex]
AGEL 6 & DESJ2137-0129 \\ [1.1ex]
AGEL 7 & DESJ0537-4647 \\ [1.1ex]
AGEL 8 & DESJ2014-5757 \\ [1.1ex]
AGEL 9 & DCLS1427-0645 \\ [1.1ex]
AGEL 10 & DESJ2335-5152 \\ [1.1ex]
AGEL 11 & PS1J2336 \\ [1.1ex]
AGEL 12 & DESJ2334-6404 \\ [1.1ex]
\bottomrule
\bottomrule
\end{tabular}
\caption[Target names]{Mapping between target ids of the observational data samples and the labels used in this paper.}
\label{tab:targetnames}
\end{table}

\end{document}